# Progressive failure simulation of composite materials using the anisotropic phase field method

Yuanfeng Yu[1], Xiaoya Zheng[2], Peng Li[3], Jinyou Xiao[2]

[1] *School of Aeronautics, Northwestern Polytechnical University, Xi'an 710072, China*

[2] *School of Astronautics, Northwestern Polytechnical University, Xi'an 710072, China*

[3] *Xi'an Modern Control Technology Research Institute, Xi'an 710065, China*

## Abstract

An effective computational framework of an anisotropic phase field model is used to explore the effects of interface properties, effective critical energy release rates and hole shapes on the failure process of fibre-reinforced composites in this paper. In this framework, the phase field method is solved under the background of the finite element method, and the influence of strong and weak interfaces on the failure strength of composite materials is studied. The results show that when the material fails, the strong interface shows brittleness, and the weak interface shows toughness, which can improve the failure strength of the material. At the same time, the effective critical energy release rate is introduced in the calculation, which reduces the fracture toughness of the material, makes the prediction results more consistent with the experimental results, and improves the accuracy of the results. The effect of hole shapes on the failure of the composites is also explored, and the crack propagation of different hole shapes is captured, which shows that the bearing capacity of a plate with a hole is not only related to the shape of the hole but also related to the location of the hole concentrated stress and the number of locations where the hole bears the load. The bearing capacity of the material can be improved by changing the shape of the hole. These results reveal the influence of different factors on the failure of composites and lay a foundation for the effective design of composites.

**Keywords:** Composite material; Phase field method; Interface; Crack propagation; Progressive failure

# 1 Introduction

Composite materials are widely used in aerospace, automobile, power and other fields because of their high specific strength, high specific stiffness and designability. In the service process of composite components, material damage, crack propagation, fracture, delamination, penetration and other discontinuous problems seriously threaten the safety of composite components. However, composite materials are sensitive to damage, which will reduce the residual strength of composite plates. Therefore, the establishment of a basic damage behaviour analysis and prediction model is an important prerequisite for its application in structural design. At present, the damage and failure analysis of composite materials is not perfect in both experimental and theoretical research studies, which limits their application to a certain extent. To make more effective use of the performance advantages of composite materials and give full play to the potential of composite materials, it is of great significance to further study the failure mechanism of composite structures [1].

Composite materials are typical anisotropic materials, and progressive failure analysis of structures is always difficult and challenging in engineering. At present, there are different numerical methods to simulate the failure process of composites. Among them, the multiscale method is an effective calculation method [2-4] in which the periodic representative volume element (RVE) of composite materials is first established [5, 6] and the mechanical properties are calculated at meso-scale. Then, the macroscopic-mesoscopic bridging matrix is built according to the homogenization method and continuous damage mechanics (CDM). Through the bridging mechanism, the mechanical properties of the meso-component materials are transferred to the macro-materials to analyse the damage process of composite materials. Although the calculation accuracy of this method is high, the efficiency is low; thus, the damage problem of a large model cannot be solved. Therefore, reduced-order modelling (ROM) is introduced to improve the computational efficiency [7-9]. The cohesive element method is widely used to know the potential failure path in advance, such as delamination between different layers [10-12]. However, the potential propagation path of matrix cracks depends on the in situ stress state and is unknown a priori. It is impractical to insert cohesive elements into the calculation model everywhere to capture any possible matrix cracks. The interaction between matrix cracking and delamination may further complicate modelling tasks, such as delamination migration [13-15]. To model the fracture problem of unknown crack paths conveniently and effectively, several other numerical methods have been introduced, such as the extended finite element method (XFEM) [16, 17], phantom node method [18, 19], and augmented finite element method [20, 21]. The XFEM provides an effective and convenient way to represent the evolution of cracks on the grid.

Recently, an effective analysis method called the phase field method (PFM) has emerged in the field of computation and has been rapidly developed and widely studied. The PFM overcomes the inability of the classical Griffith theory to address crack initiation, bifurcation and combination. At the same time, it also does not require additional damage criteria and crack tracking criteria [22]. To date, the PFM has been applied to many research fields, such as brittle fracture [23, 24], dynamic fracture [25, 26], cohesive fracture [27, 28], and ductile fracture [29, 30]. Compared with the classical progressive damage method, the PFM avoids the use of phenomenological criteria and requires fewer material parameters. Recently, some new phase field models have been developed to solve the problem of composite failure [31-34]. For example, Alessi and Freddi [35] used the phase field model to study the failure process of unidirectional hybrid laminates. Bleyer and Alessi [36] proposed a longitudinal/transverse damage model to simulate the longitudinal and transverse failure of composites. Quintanas-Corominas et al. [37] proposed a new phase field model with the Gibbs free energy to simulate the interlaminar and intralaminar failure of fibre-reinforced composites. Hirshikesh et al. [38] studied crack propagation in variable stiffness composites using the phase field model. In [39], Zhang et al. used a new three-dimensional crack surface density function phase field model to study the fracture of anisotropic composite laminates, and in [40], Zhang et al. used a mixed phase field model combined with cohesive elements to study the delamination and debonding of cross ply composite laminates. In [41], Zhang et al. derived a new criterion mathematically and embedded the widely used Hashin criterion for matrix crack initiation into the phase field model to accurately predict the failure mode of composite materials.

Due to the special microstructure of composites, when the composites are damaged, many factors are involved and coupled, and the mechanical properties are often affected by the interface properties, fibre direction and critical energy release rate. At the same time, in the process of using composite materials, holes are often made in the composite structure to connect different parts. When there are holes in the structure, the mechanical properties will change. Due to the different stress concentrations caused by different holes, the final mechanical properties will change differently due to the hole shapes. To understand the influence of the interface, fibre direction, and hole shape on the failure of composite materials, the following research was carried out in this study:

    (1) The PFM is used to study the strong and weak interface properties of anisotropic composite materials to understand the influence of different interfaces on the failure stress of the material.

    (2) The effective critical energy release rate on the failure stress of the composite material is analysed.

    (3) The failure stress and the damage evolution process of composite materials with different holes is analysed.

After an introduction, this paper is arranged as follows: Section 2 introduces the anisotropic phase field theory. The implementation of the phase field model is described in Section 3. The phase field model that is used to study the influence of different factors on the failure strength of composite materials is discussed in Section 4. Finally, the conclusion is summarized in Section 5.

## 2 Anisotropic phase field model

In the phase field model, the topology of a sharp crack is approximated to a diffuse crack by the finite length, which ensures that the sharp crack has a finite width. Consequently, some functions can be used to represent the approximate diffuse crack to characterize the damage and fracture process of the material.

**2.1 Fracture energy functional**

2.1.1 Crack dissipative functional

For a continuous solid, a sharp crack can be approximated by a diffuse phase field distribution, where $\Gamma$ represents the completely destroyed crack surface, as shown in Fig. 1. An auxiliary variable $d(x) \in [0,1]$ can be introduced to describe the sharp crack topology. When $d=0$, the intact state of the material is represented, and when $d=1$, the complete fracture of the material is indicated.

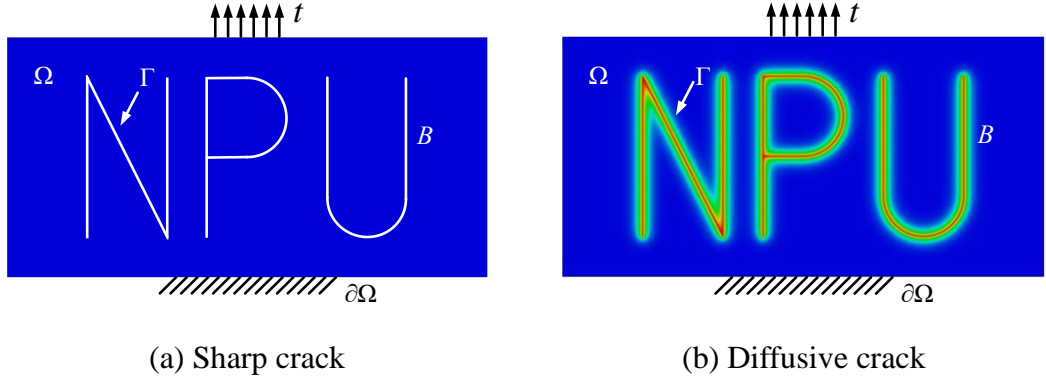

(a) Sharp crack  (b) Diffusive crack

**Fig. 1. Crack topology**

Here, an exponential function can be used to approximate the nonsmooth crack phase field [42],

$$d(x) = e^{-|x|/l} \tag{1}$$

so that the sharp cracks are diffused, representing a regularized crack topology, as shown in Fig. 1(b). The parameter $l$ controls the width of the dispersed crack to characterize the dispersion degree of the crack. In this case, we can use functional

$$\Gamma(d) = \frac{1}{2l} \int_\Omega \left\{ d^2 + l^2 |\nabla d|^2 \right\} dV \tag{2}$$

to characterize the crack surface topology and define the crack surface density function $\gamma$ on the unit area (or length) as

$$\gamma(d, \nabla d) = \frac{1}{2l} \left( d^2 + l^2 |\nabla d|^2 \right). \tag{3}$$

When Eq. (3) is used to approximate the crack topology, the influence of the material direction on the crack propagation direction is not considered. While the PFM is used to study the damage of anisotropic composites, especially for the study of anisotropic fibre-reinforced composites, it is necessary to consider the influence of fibre orientation on the damage results, as shown in Fig. 2, where $\theta$ indicates the fibre direction.

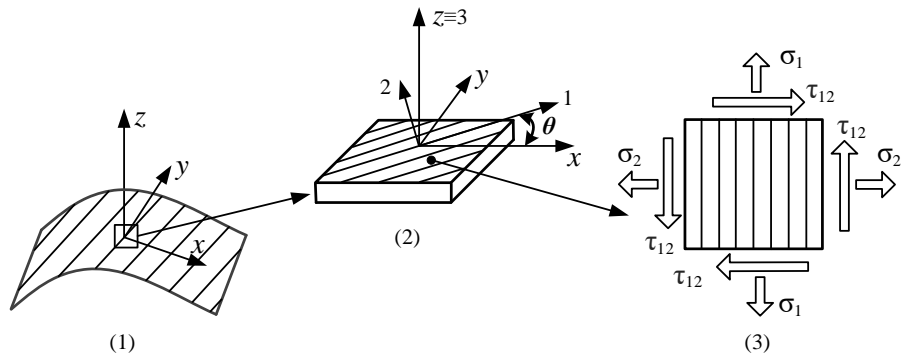

**Fig. 2. Composite material model: (1) global coordinate system; (2) local coordinate system; and (3) stress in the local coordinate system**

To simulate the damage process of anisotropic composites by the PFM, the structure tensor $\mathbf{A}$ should be introduced [43]. Therefore, the geometric crack function $\gamma(d, \nabla d)$ in Eq. (3) will become $\gamma(d, \nabla d, \theta)$, as shown in the following equation

$$\gamma(d, \nabla d, \theta) = \frac{1}{2l}\left(d^2 + l^2 \nabla d \cdot \mathbf{A} \cdot \nabla d\right), \tag{4}$$

For the structure tensor $\mathbf{A}$, there are currently several different choices [37, 44-46].

$$\begin{aligned} \mathbf{A}_1 &:= \mathbf{I} + \alpha \mathbf{M}; \\ \mathbf{A}_2 &:= \mathbf{I} + \alpha(\mathbf{I} - \mathbf{N}); \\ \mathbf{A}_3 &:= \mathbf{M} + \alpha(\mathbf{I} - \mathbf{M}). \end{aligned} \tag{5}$$

The structure tensor can describe the effect of fibre direction on the damage. In Eq. (5), $\mathbf{I}$ can be used to express the influence of an isotropic matrix on damage, and $\mathbf{M}$ and $\mathbf{N}$ can be used to express the influence of fibres on damage. Combining them can express the influence of fibre direction on crack propagation in anisotropic composites. Here, $\mathbf{I} := \delta_{ij} e_i \otimes e_j$ is the second-order structure tensor, where $\delta_{ij}$ denotes the Kronecker delta, and the tensors $\mathbf{M}$ and $\mathbf{N}$ are

$$\begin{aligned} \mathbf{M} &:= a \otimes a, \\ \mathbf{N} &:= b \otimes b, \end{aligned} \tag{6}$$

where $a$ is the fibre direction vector and $b$ is the unit vector normal to the fibre direction.

In the two-dimensional case, the calculation results of the structure tensors $\mathbf{A}_1$ and $\mathbf{A}_2$ in the Eq.(5) are the same, but in the three-dimensional case, the calculation results of them are different. Substituting in $\mathbf{A}_1$ in Eq.(5) into Eq.(4), the anisotropic crack function is obtained

$$\begin{aligned} \gamma(d, \nabla d, \theta) &= \frac{1}{2l}\left(d^2 + l^2 \nabla d \cdot (\mathbf{I} + \alpha \mathbf{M}) \cdot \nabla d\right) = \\ &\quad \frac{1}{2l}\left(d^2 + l^2 \nabla d \cdot \nabla d + \alpha l^2 (\nabla d \cdot a)^2\right), \end{aligned} \tag{7}$$

It can be seen from Eq.(7) that the parameter $\alpha$ can be used as a weighting coefficient of the anisotropic crack model to punish damage in non-fiber directions. At that time, $\alpha \to \infty$, and $\nabla d \cdot a = 0$, which means that the crack will be parallel to the material direction; in other words, the crack will grow along the material direction. For some special cases, it can be obtained that

$$\gamma(d, \nabla d, \theta) = \begin{cases} \dfrac{1}{2l}\left(d^2 + l^2 |\nabla d|^2 (1+\alpha)\right) & \nabla d \parallel \boldsymbol{a} \\ \dfrac{1}{2l}\left(d^2 + l^2 |\nabla d|^2\right) & \nabla d \perp \boldsymbol{a} \end{cases} \tag{8}$$

It can be seen from Eq.(8) that to make the crack $(d=1)$ propagate along the material direction, this equation must satisfy

$$0 < 1 + \alpha, \tag{9}$$

Therefore, we obtain

$$-1 < \alpha < \infty. \tag{10}$$

The structure tensor $\mathbf{A}_3$ in Eq.(5) can be re-expressed as

$$\mathbf{A}_3 := \alpha \mathbf{I} + (1-\alpha)\mathbf{M} \tag{11}$$

At this time, substituting Eq.(11) into Eq.(4), the anisotropic crack function is obtained

$$\begin{aligned}\gamma(d, \nabla d, \theta) &= \frac{1}{2l}\left(d^2 + l^2 \nabla d \cdot (\alpha \mathbf{I} + (1-\alpha)\mathbf{M}) \cdot \nabla d\right) = \\ &\frac{1}{2l}\left(d^2 + \alpha l^2 \nabla d \cdot \nabla d + (1-\alpha) l^2 (\nabla d \cdot \boldsymbol{a})^2\right),\end{aligned} \tag{12}$$

In the same way, for some special cases, it can be obtained that

$$\gamma(d, \nabla d, \theta) = \begin{cases} \dfrac{1}{2l}\left(d^2 + l^2 |\nabla d|^2\right) & \nabla d \parallel \boldsymbol{a} \\ \dfrac{1}{2l}\left(d^2 + \alpha l^2 |\nabla d|^2\right) & \nabla d \perp \boldsymbol{a} \end{cases} \tag{13}$$

It can be seen from Eq.(11) and (13) that in order to make the crack propagate along the material direction, the equation needs to satisfy

$$0 < \alpha < 1, \tag{14}$$

It can be seen from Eq.(12) and (13) that when $\alpha \to 0$, $\nabla d \cdot \boldsymbol{a} = 0$ can be obtained. At this time, the crack will be parallel to the fiber direction, and the crack will expand along the fiber direction, but in this way, the ability of $\alpha$ to punish damage in the non-fiber direction is weakened. Therefore, compared with the structure tensor $\mathbf{A}_2$ and $\mathbf{A}_2$, the penalty effect of the structure tensor $\mathbf{A}_3$ in different fiber angles will be affected. Therefore, structure tensor $\mathbf{A}_1$ is chosen in this article.

$$\mathbf{A} = \mathbf{A}_1 := \mathbf{I} + \alpha \mathbf{M}. \tag{15}$$

Regarding the value of the parameter $\alpha$, in [44], Teichtmeister et al. analysed some characteristics of $\alpha$. In [39], Zhang et al. gave the specific value of $\alpha$ and proved it, but it can only be applied to one-dimensional situations.

When the parameter $\alpha$ in Eq.(15) is equal to 0, $\alpha = 0$, the structure tensor is $\mathbf{A} = \mathbf{I}$, and the model will be restored to the isotropic model (3). At this time, the damage of the material in different directions is the same. While the parameter $\alpha$ is not equal to 0, $\alpha \neq 0$, it can be used to simulate the damage of anisotropic composite materials. For example, if the fibre direction is 0°, namely, $\boldsymbol{e}_1 = \boldsymbol{a} = [1,0]$, then damage will evolve along the $\boldsymbol{e}_1$ direction; if the fibre direction is 45°, namely, $\boldsymbol{n} = \boldsymbol{a} = \left[\sqrt{2}/2, \sqrt{2}/2\right]$, then damage will evolve along the $\boldsymbol{n}$ direction, as shown in Fig. 3.

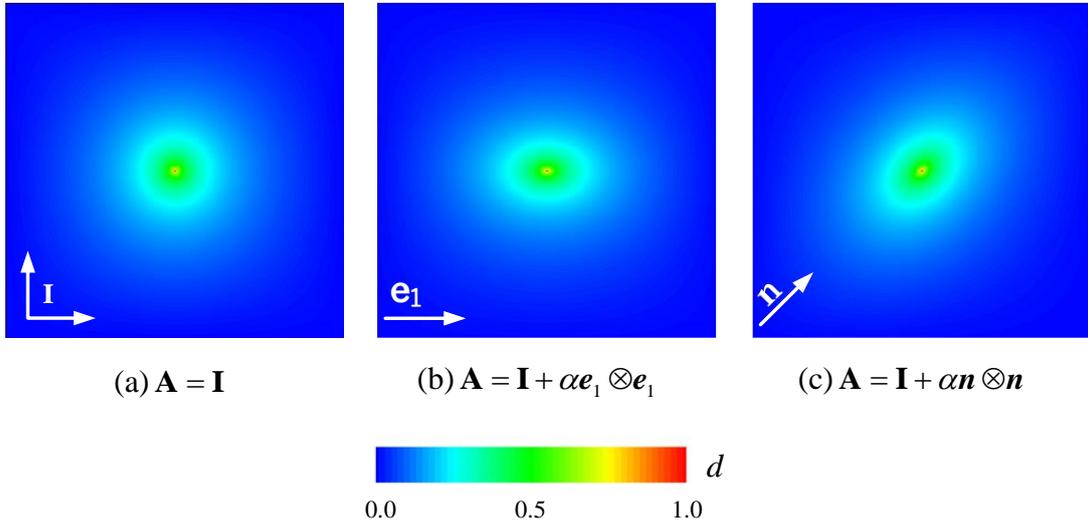

(a) $\mathbf{A} = \mathbf{I}$  (b) $\mathbf{A} = \mathbf{I} + \alpha \boldsymbol{e}_1 \otimes \boldsymbol{e}_1$  (c) $\mathbf{A} = \mathbf{I} + \alpha \boldsymbol{n} \otimes \boldsymbol{n}$

**Fig. 3. Damage evolution results**

According to the fracture surface function $\Gamma(d)$ introduced in Eq. (2), the work required to generate a new crack topology can be defined

$$\Pi_c(d,\theta) = \int_\Omega G_c \gamma(d, \nabla d, \theta) dV$$
$$= \int_\Omega \frac{G_c}{2l} \left(d^2 + l^2 \nabla d \cdot \mathbf{A} \cdot \nabla d\right) dV. \tag{16}$$

where $\gamma(d, \nabla d, \theta)$ is the crack surface density function defined in Eq. (4), $G_c$ is the critical energy release rate of the Griffith type and is the work required to generate a new crack unit length [47].

2.1.2 Strain energy functional

According to the theory of linear elasticity, the global energy storage functional

can be expressed as

$$\Pi_e(\boldsymbol{u},d,\theta) = \int_B \psi(\varepsilon(\boldsymbol{u}),d,\theta)dV, \tag{17}$$

In the two-dimensional (2D) case, the stress and strain in Eq. (17) are

$$\begin{bmatrix} \sigma_x \\ \sigma_y \\ \tau_{xy} \end{bmatrix} = [\boldsymbol{T}] \begin{bmatrix} \sigma_1 \\ \sigma_2 \\ \tau_{12} \end{bmatrix}, \begin{bmatrix} \varepsilon_x \\ \varepsilon_y \\ \gamma_{xy} \end{bmatrix} = [\boldsymbol{T}^{-1}]^{\mathrm{T}} \begin{bmatrix} \varepsilon_1 \\ \varepsilon_2 \\ \gamma_{12} \end{bmatrix}. \tag{18}$$

where $[\sigma_x,\sigma_y,\tau_{xy},\varepsilon_x,\varepsilon_y,\gamma_{xy}]$ is the stress and strain in the global coordinate system and $[\sigma_1,\sigma_2,\tau_{12},\varepsilon_1,\varepsilon_2,\gamma_{12}]$ is the stress and strain in the local coordinate system. $\boldsymbol{T}$ is the coordinate transformation matrix, and its expression is

$$\boldsymbol{T} = \begin{bmatrix} \cos^2\theta & \sin^2\theta & -2\sin\theta\cos\theta \\ \sin^2\theta & \cos^2\theta & 2\sin\theta\cos\theta \\ \sin\theta\cos\theta & -\sin\theta\cos\theta & \cos^2\theta - \sin^2\theta \end{bmatrix}. \tag{19}$$

According to Eq. (18) and the stress-strain relationship in the local coordinate system

$$\begin{bmatrix} \sigma_1 \\ \sigma_2 \\ \tau_{12} \end{bmatrix} = [\boldsymbol{E}_0] \begin{bmatrix} \varepsilon_1 \\ \varepsilon_2 \\ \gamma_{12} \end{bmatrix}. \tag{20}$$

The relationship between stress and strain in the global coordinate system can be obtained

$$\begin{bmatrix} \sigma_x \\ \sigma_y \\ \tau_{xy} \end{bmatrix} = [\boldsymbol{T}][\boldsymbol{E}_0][\boldsymbol{T}]^{\mathrm{T}} \begin{bmatrix} \varepsilon_x \\ \varepsilon_y \\ \gamma_{xy} \end{bmatrix}. \tag{21}$$

where $\boldsymbol{E}_0$ is the undamaged elastic stiffness matrix, and its expression is

$$\boldsymbol{E}_0 = \begin{bmatrix} E_{11} & E_{12} & 0 \\ E_{12} & E_{22} & 0 \\ 0 & 0 & E_{66} \end{bmatrix}. \tag{22}$$

In Eq. (22), the parameters are

$$E_{11} = \frac{E_1}{1-v_{12}v_{21}}, E_{22} = \frac{E_2}{1-v_{12}v_{21}}$$

$$E_{12} = \frac{v_{21}E_1}{1-v_{12}v_{21}} = \frac{v_{12}E_2}{1-v_{12}v_{21}}, E_{66} = G_{12} \tag{23}$$

To avoid damage evolution in compression, the strain energy $\psi(\varepsilon,d,\theta)$ is

decomposed into two parts:

$$\psi(\varepsilon,d,\theta) = g(d)\psi_0^+ + \psi_0^- = \frac{1}{2}g(d)(\varepsilon^+)^T E_0 \varepsilon^+ + \frac{1}{2}(\varepsilon^-)^T E_0 \varepsilon^- \quad (24)$$

where $g(d)$ is the degradation function, $g(d) = (1-d)^2$, and the parameter $k$ is the residual stiffness coefficient to ensure numerical stability. For two-dimensional problems, $\varepsilon^+$ and $\varepsilon^-$ are

$$\varepsilon^+ = \begin{Bmatrix} \langle\varepsilon_{11}\rangle_+ \\ \langle\varepsilon_{22}\rangle_+ \\ 2\varepsilon_{12} \end{Bmatrix}, \quad \varepsilon^- = \begin{Bmatrix} \langle\varepsilon_{11}\rangle_- \\ \langle\varepsilon_{22}\rangle_- \\ 0 \end{Bmatrix} \quad (25)$$

Similarly, for three-dimensional problems, $\varepsilon^+$ and $\varepsilon^-$ are

$$\varepsilon^+ = \{\langle\varepsilon_{11}\rangle_+, \langle\varepsilon_{22}\rangle_+, \langle\varepsilon_{33}\rangle_+, 2\varepsilon_{12}, 2\varepsilon_{13}, 2\varepsilon_{23}\}^T,$$
$$\varepsilon^- = \{\langle\varepsilon_{11}\rangle_-, \langle\varepsilon_{22}\rangle_-, \langle\varepsilon_{33}\rangle_-, 0, 0, 0\}^T. \quad (26)$$

Here, $\langle x \rangle_\pm = (x \pm |x|)/2$ is the bracket operator. Therefore, the corresponding energy functional $\psi_0^+$ can be expressed as

$$\psi_0^+ = \psi_{TEEDF}^+ + \psi_{TEEDM}^+ + \psi_{TSEDM}^+ \quad (27)$$

where $\psi_{TEEDF}^+$, $\psi_{TEEDM}^+$ and $\psi_{TSEDM}^+$ represent the tensile energy density of the fibre, the tensile energy density of the matrix and the transverse shear energy density of the matrix, respectively. The respective specific expression forms are

$$\psi_{TEEDF}^+ = \frac{1}{2}\langle\varepsilon_{11}\rangle_+ \langle\sigma_{11}\rangle_+, \quad 2D \text{ and } 3D \quad (28)$$

$$\psi_{TEEDM}^+ = \begin{cases} \frac{1}{2}\langle\varepsilon_{22}\rangle_+ \langle\sigma_{22}\rangle_+, & 2D \\ \frac{1}{2}\langle\varepsilon_{22}\rangle_+ \langle\sigma_{22}\rangle_+ + \frac{1}{2}\langle\varepsilon_{33}\rangle_+ \langle\sigma_{33}\rangle_+, & 3D \end{cases} \quad (29)$$

$$\psi_{TASDEM}^+ = \begin{cases} \frac{1}{2}\tau_{12}\gamma_{12}, & 2D \\ \frac{1}{2}\tau_{12}\gamma_{12} + \frac{1}{2}\tau_{23}\gamma_{23} + \frac{1}{2}\tau_{13}\gamma_{13}, & 3D \end{cases} \quad (30)$$

Similarly, $\psi_0^-$ can be expressed as

$$\psi_0^- = \psi_{CEDF}^- + \psi_{CEDM}^- \quad (31)$$

where $\psi_{CEDF}^-$ and $\psi_{CEDM}^-$ represent the compressive energy density of the fibre and the compressive energy density of the matrix, respectively, and the specific forms are

$$\bar{\psi}_{CEDF} = \frac{1}{2}\langle\varepsilon_{11}\rangle_{-}\langle\sigma_{11}\rangle_{-}, \quad 2D \text{ and } 3D \tag{32}$$

$$\bar{\psi}_{CEDM} = \begin{cases} \frac{1}{2}\langle\varepsilon_{22}\rangle_{-}\langle\sigma_{22}\rangle_{-}, & 2D \\ \frac{1}{2}\langle\varepsilon_{22}\rangle_{-}\langle\sigma_{22}\rangle_{-} + \frac{1}{2}\langle\varepsilon_{33}\rangle_{-}\langle\sigma_{33}\rangle_{-}, & 3D \end{cases} \tag{33}$$

## 2.2 Governing equations of the phase field model

According to the crack dissipation functional (16) and energy storage functional (17), the total energy functional is introduced

$$\Pi(\boldsymbol{u},d,\theta) = \Pi_e(\boldsymbol{u},d,\theta) + \Pi_c(d,\theta) - \Pi_{ext}(\boldsymbol{u}) \tag{34}$$

Here, the external energy functional $\Pi_{ext}(\boldsymbol{u})$ is defined as

$$\Pi_{ext}(\boldsymbol{u}) = \int_{\mathcal{B}} \boldsymbol{b}\cdot\boldsymbol{u}\,dV + \int_{\partial\mathcal{B}} \boldsymbol{t}\cdot\boldsymbol{u}\,dA. \tag{35}$$

where $b$ is the volume force in region $\mathcal{B}$ and $t$ is the external force on surface $\partial\mathcal{B}$. According to the boundary condition of the total energy functional

$$\begin{cases} \delta\Pi = 0, & \delta d > 0 \\ \delta\Pi > 0, & \delta d = 0 \end{cases} \tag{36}$$

The following strong mechanical equilibrium equations and phase field evolution equations can be obtained:

$$\begin{aligned} &\nabla\cdot\boldsymbol{\sigma} + b = 0, \quad in\ \mathcal{B} \\ &\frac{G_c}{l}d + G_c l \nabla d\cdot\boldsymbol{A}\cdot\nabla d - 2(1-d)\mathcal{H} \geq 0, \quad in\ \Omega \end{aligned} \tag{37}$$

In Eq. (37), $\mathcal{H}(\boldsymbol{x},t) := \max_{s\in[0,t]}\left[\psi_0^+(\varepsilon(\boldsymbol{x},t))\right]$, and $\psi_0^+(\varepsilon(\boldsymbol{x},t))$ is as shown in Eq. (27). The boundary condition is

$$\begin{aligned} &\boldsymbol{\sigma}\cdot\boldsymbol{n} = \boldsymbol{t}, \quad on\ \partial\mathcal{B} \\ &\frac{\partial\gamma}{\partial\nabla d}\cdot\boldsymbol{n}_{\Omega} = 0, \quad on\ \partial\Omega \end{aligned} \tag{38}$$

Here, to facilitate the analysis of the stress characteristics at failure, it is assumed that there is no compressive strain in the material; that is, $\psi_0^- = 0$, $\psi_0 = \frac{1}{2}E_0\varepsilon^2$, and the stress is $\sigma = (1-d)^2 E_0\varepsilon$.

When there is a crack surface gradient change in the model, $\nabla d \neq 0$, it can be

deduced from Eq. (37)$_2$

$$\sigma = \sqrt{(1-d)^3 \left[ \frac{E_0 G_c}{l} d + E_0 G_c l \nabla d \cdot \mathbf{A} \cdot \nabla d \right]}$$

$$= \sqrt{(1-d)^3 \left[ \frac{E_0 G_c}{l} d + E_0 G_c l \nabla d \cdot (\mathbf{I} + \alpha \mathbf{M}) \cdot \nabla d \right]} \tag{39}$$

It can be seen from Eq. (39) that when the damage field $d$ is fixed, the stress $\sigma$ will increase with increasing parameter $\alpha$, and vice versa. Since it is usually set to $\alpha > 0$, the result will be larger. At the same time, we find from Eq. (39)

$$\sigma \propto \sqrt{G_c} \tag{40}$$

The stress $\sigma$ increases with increasing energy release rate $G_c$ and decreases with decreasing energy release rate $G_c$. To reduce the influence of the parameters $\alpha$ on the stress results and to eliminate the influence of the mesh size on the results, the effective critical energy release rate is defined as

$$G_c^e = G_c \left(1 - \frac{l_m}{l}\right) f(\alpha) \tag{41}$$

where $l_m$ is the minimum element size and $f(\alpha)$ is a positive function of the parameter $\alpha$, that is, $f(\alpha) > 0$. In addition, $f(\alpha)$ should meet the following requirements:

$$\begin{cases} f(\alpha) = 1, & \alpha = 0 \\ f(\alpha) < 1, & \alpha > 0 \\ f(\alpha) \to \varepsilon, & \alpha \gg 1 \end{cases} \tag{42}$$

Here, $0 < \varepsilon < 1$ is a finite value; usually, the value of $\alpha$ will not be very large. To render a simple calculation, $f(\alpha)$ is taken as

$$f(\alpha) = 1 - \frac{\alpha}{10^3} \tag{43}$$

Therefore, the effective critical energy release rate is

$$G_c^e = G_c \left(1 - \frac{l_m}{l}\right)\left(1 - \frac{\alpha}{10^3}\right) \tag{44}$$

## 3 Implementation of the phase field model

At present, there are many different methods to solve the governing equations of the phase field model, such as the fast Fourier transform (FFT) [24], scaled boundary finite element method (SBFEM) [48], and material point method (MPM) [49]. Although these methods are efficient in terms of the calculation, they have not been applied to general finite element software, which limits the application of these methods. The commercial finite element software ABAQUS has excellent secondary development ability and can be used to solve complex model problems. Therefore, this section will study the implementation process of the phase field model in the finite element solution platform ABAQUS.

A large number of user subroutines are provided in ABAQUS, and these subroutines can help users define their own mathematical models and algorithms and make their applications more flexible and convenient. In the phase field fracture model, the user-defined element (UEL) [50] is needed because the phase field evolution equation cannot be expressed in the standard finite element solution format. The calculation flow of the UEL module in ABAQUS is shown in Fig. 4 [51].

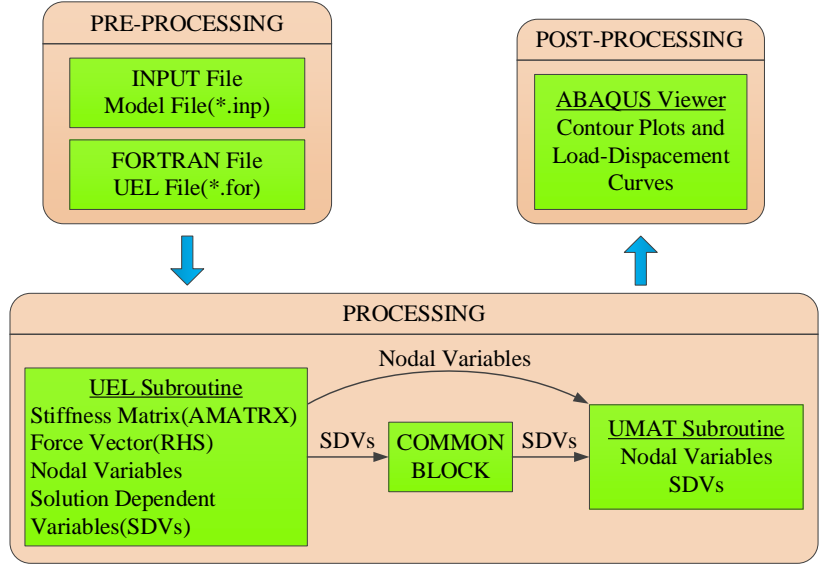

**Fig. 4. The flow of solving UEL in ABAQUS**

### 3.1 Element discretization

To solve the phase field model in ABAQUS, the variables need to be discretized first, and the displacement field $u$ and its corresponding strain field $\varepsilon$ can be discretized as

$$u(x) = \sum_{i=1}^{n} N_i u_i = Nu, \quad \varepsilon(x) = \sum_{i=1}^{n} B_i u_i = Bu, \tag{45}$$

where $N := [N_1, \cdots N_i, \cdots, N_n]$ is the interpolation matrix,

$B := [B_1, \cdots, B_i, \cdots, B_n]$ is the geometric matrix, and $N_i$ and $B_i$ in the 2D case are

$$N_i = \begin{bmatrix} N_i & 0 \\ 0 & N_i \end{bmatrix}, B_i = \begin{bmatrix} N_{i,x} & 0 \\ 0 & N_{i,y} \\ N_{i,y} & N_{i,x} \end{bmatrix} \tag{46}$$

Here, $N_i$ represents the shape function at the element node $i$, and $u_i = [u_x, u_y]_i^T$ or $u_i = [u_x, u_y, u_z]_i^T$ represents the displacement value at the element node $i$.

Similarly, the phase field $d$ and its corresponding gradient field $\nabla d$ can be discretized as

$$d(x) = \sum_{i=1}^n \bar{N}_i d_i = \bar{N}d, \ \nabla d(x) = \sum_{i=1}^n \bar{B}_i d_i = \bar{B}d. \tag{47}$$

where $\bar{N} := [\bar{N}_1, \cdots \bar{N}_i, \cdots, \bar{N}_n]$ is the interpolation matrix, $\bar{B} := [\bar{B}_1, \cdots, \bar{B}_i, \cdots, \bar{B}_n]$ is the geometric matrix, and $\bar{N}_i$ and $\bar{B}_i$ in the 2D case are

$$\bar{N}_i = N_i, \bar{B}_i = \begin{bmatrix} N_{i,x} \\ N_{i,y} \end{bmatrix} \tag{48}$$

The function variation and derivative of the corresponding displacement field and phase field are

$$\delta u(x) = \sum_{i=1}^n N_i \delta u_i = N \delta u, \ \delta \varepsilon(x) = \sum_{i=1}^n B_i \delta u_i = B \delta u$$
$$\delta d(x) = \sum_{i=1}^n \bar{N}_i \delta d_i = \bar{N} \delta d, \ \nabla \delta d(x) = \sum_{i=1}^n \bar{B}_i \delta d_i = \bar{B} \delta d \tag{49}$$

Therefore, the displacement residual flux of the whole equation system at node $i$ can be obtained

$$r_i^u(u + \Delta u, d) = f_{iint}^u - f_{iext}^u =$$
$$\int_\mathcal{B} (B_i)^T \sigma dV - \left( \int_\mathcal{B} (N_i)^T b dV + \int_{\partial \mathcal{B}} (N_i)^T t dA \right) \tag{50}$$

The corresponding phase field residual flux is

$$r_i^d(u, d + \Delta d) = \int_\Omega \left\{ \left[ \frac{G_c}{l} d - 2(1-d)\mathcal{H} \right] (\bar{N}_i)^T + G_c l (\bar{B}_i)^T \cdot \mathbf{A} \cdot \nabla d \right\} dV \tag{51}$$

By solving the residual fluxes $r_i^u(u + \Delta u, d) = 0$ and $r_i^d(u, d + \Delta d) \geq 0$ defined above, the corresponding displacement field and phase field problems can be obtained.

Here, staggered schemes are used to solve the governing equations of the above phase field model [52, 53], which is a semi-implicit solution method with robustness and good convergence. The structure of the staggered schemes in ABAQUS is shown in Fig. 5.

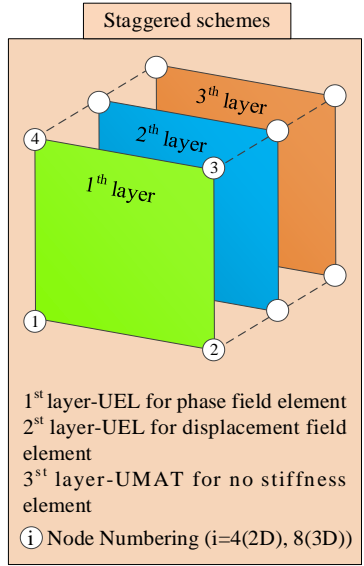

Fig. 5. The structure of staggered schemes in ABAQUS

## 3.2 Example verification

To illustrate the feasibility of the anisotropic phase field model used in this paper to analyse the damage of composite materials, an example is used in this section. The material properties of the model in the example are shown in Table 1. The model is composed of unidirectional composite plates with different fibre directions. Here, two different fibre direction combinations are given: one fibre direction is {0°/-22.5°/0°/22.5°}, and the other is {0°/45°/-45°/90°}. There is a V-shaped crack in the middle of the model. The length and width of the model are 1 mm, and the load is 0.05 mm. The geometric model and boundary conditions are shown in Fig. 6, $\theta$ is the fibre direction, and the simulation results are shown in Fig. 7.

**Table 1 Material parameters**

| Material | Parameter | Value | Unit |
| --- | --- | --- | --- |
|  | $E_{11}$ | 171 | GPa |
| Property | $E_{22}$ | 9.08 | GPa |
|  | $G_{12}$ | 5.29 | GPa |

| | | |
|---|---|---|
| $v_{12}$ | 0.32 | - |
| $Gc_{11}$ | $2.77 \times 10^{-4}$ | kN/mm |
| $Gc_{22}$ | $9.78 \times 10^{-2}$ | kN/mm |

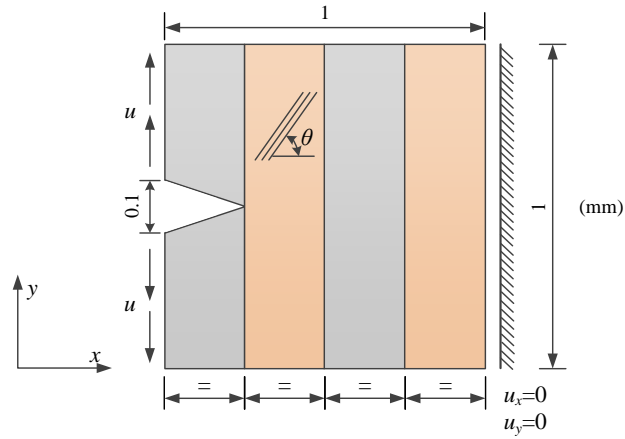

**Fig. 6. Geometric model and boundary conditions ($u$=0.05 mm)**

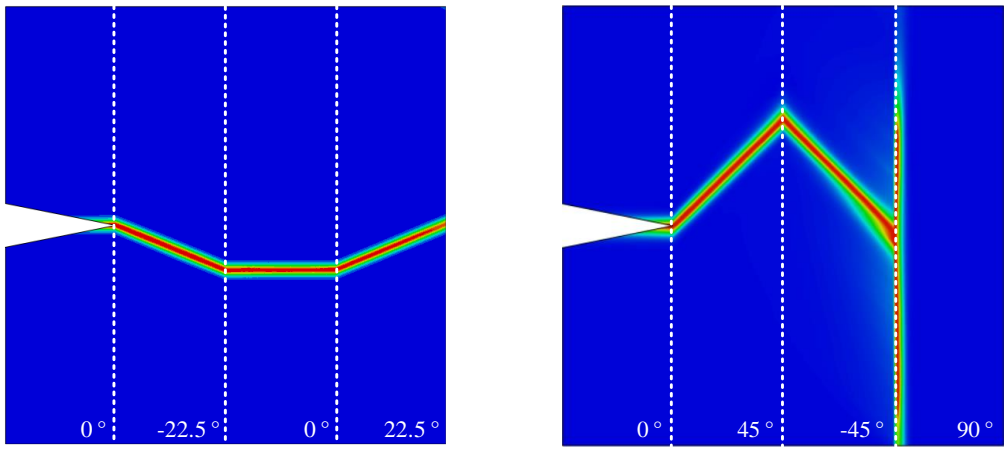

(a)          (b)

(1) Simulation results of structure tensor $\mathbf{A_1}$

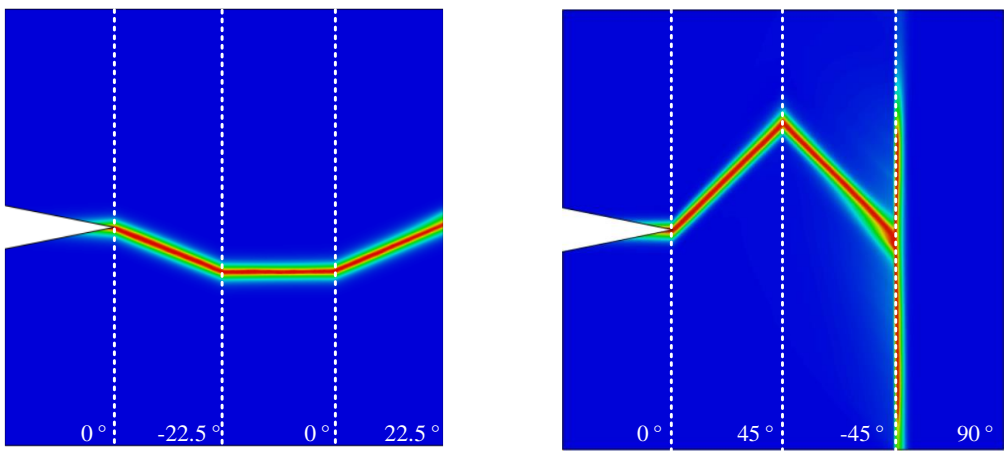

(a)          (b)

(2) Simulation results of structure tensor **A₂**

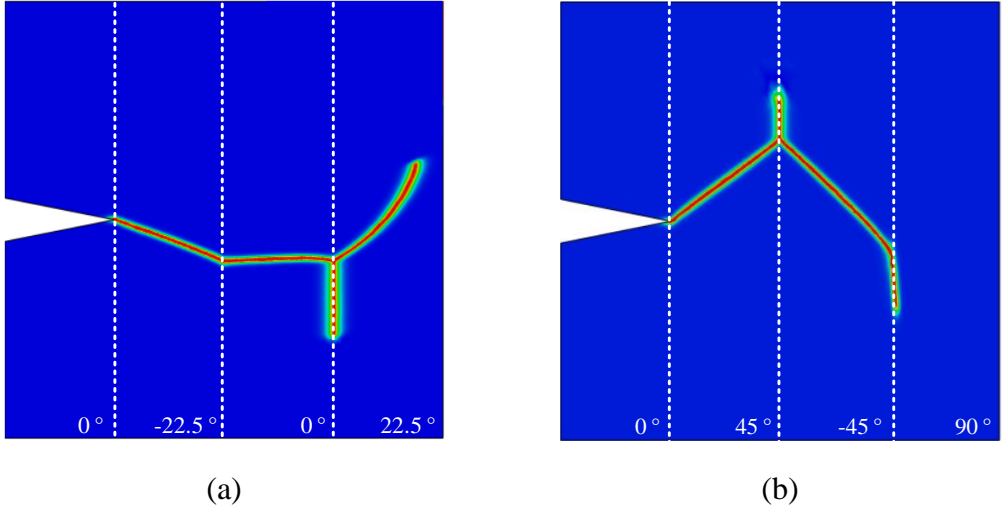

(a)          (b)

(3) Simulation results of structure tensor **A₃**

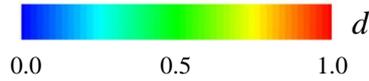

**Fig. 7. Simulation results of different structure tensors: (a) {0 °/-22.5 °/0 °/22.5 °} and (b) {0 °/45 °/-45 °/90 °}**

From the comparison results of Fig. 7(1) and (2), the simulation results of the them are consistent and consistent with the theoretical analysis results in reference [37]. The damage evolves in the direction of energy reduction, that is, the damage evolves along the direction of fibre with lower fracture toughness. It shows that when the structural tensors $A_1$ and $A_2$ are used, the calculation results are the same in the two-dimensional case. At the same time, from the simulation results in Fig. 7(1b), the phase field method has the ability to simulate crack bifurcation. When α=90°, the damage result obtained by the simulation is consistent with the fiber direction. It can be found from Figure 7(3) that the simulation result using the structure tensor $A_3$ is poor, which is quite different from the theoretical analysis result. This example verifies the accuracy of the structural tensor (15) used in this paper, and illustrates the feasibility and rationality of the anisotropic phase field model calculation framework established in this paper to simulate the damage problem of composite materials.

## 4 Results and discussion

In this section, the PFM is used to study the failure characteristics of anisotropic composites. The first example is used to discuss the failure characteristics of composites with different interface properties. The second example is used to study the influence of the effective critical energy release rate on the failure stress of composites with different fibre directions. The third example is used to investigate the influence of hole shapes on the failure of composites.

**4.1 Interfacial properties analysis**

The interface is an important part of composite materials, and it is the "bond" between the fibre and matrix. It plays an important role in the overall performance of composite materials. Because the failure of composites includes the fracture of load-bearing fibre and matrix and the complex combination of crack propagation along the interface, the fracture toughness of composites not only depends on the properties of component materials but also strongly depends on the characteristics at the interface. Different interface properties will affect the ultimate failure strength of composites, and the strong interface properties between the fibre and matrix will lead to a low fracture toughness and reduce the material's failure strength, while the weak interface strength will enhance the material's fracture toughness [54, 55]. Therefore, the design of the interfacial properties has an important influence on the ultimate strength of composites. In this section, we will study the influence of different interfacial strengths on the properties of composites.

The effects of strong and weak interfaces on the mechanical properties of materials will be studied by using a macroscopic anisotropic composite model with notches. The material parameters of the model in the example are shown in Table 2. The interface is assumed to be isotropic. The critical energy release rate Gc of the weak interface is 1/10 of the in-plane $Gc_{22}$, and the critical energy release rate Gc of the strong interface is $Gc_{22}$ and ten times that of $Gc_{22}$. The length and width of the model are both 1 mm. There is an initial crack of 0.05 mm in the middle of the model. The geometric model and boundary conditions are shown in Fig. 8. The simulation results of different interface properties are shown in Fig. 9, and the force-displacement curves are shown in Fig. 10.

**Table 2 Material parameters**

| Material | Parameter | Value | Unit |
|---|---|---|---|
| | $E_{11}$ | 171 | GPa |
| Intralaminar | $E_{22}$ | 9.08 | GPa |
| | $G_{12}$ | 5.29 | GPa |
|   | $v_{12}$ | 0.32 | - |
|   | $Gc_{11}$ | $9.78 \times 10^{-2}$ | kN/mm |
|   | $Gc_{22}$ | $2.77 \times 10^{-4}$ | kN/mm |
| Weak Interface | E | 9.08 | GPa |
|   | $v$ | 0.32 | - |
|   | Gc | $2.77 \times 10^{-5}$ | kN/mm |
| Strong Interface 1 | E | 9.08 | GPa |
|   | $v$ | 0.32 | - |
|   | Gc | $2.77 \times 10^{-4}$ | kN/mm |
| Strong Interface 2 | E | 9.08 | GPa |
|   | $v$ | 0.32 | - |
|   | Gc | $2.77 \times 10^{-3}$ | kN/mm |

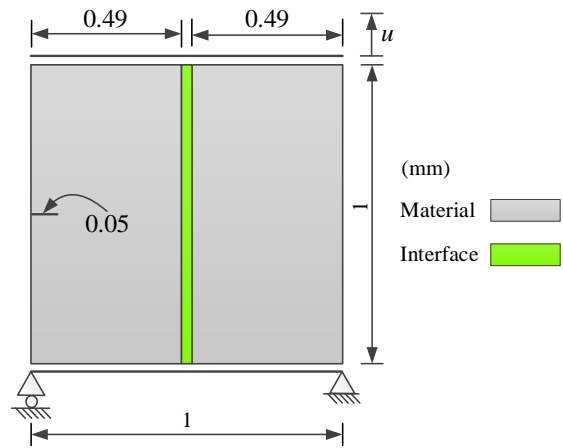

**Fig. 8. Geometric model and boundary conditions**

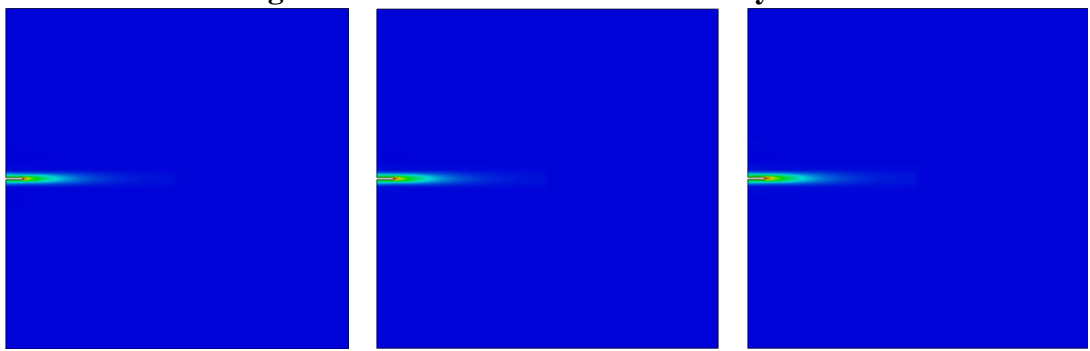

| (a) | (b) | (c) |

(1) Initial damage stage

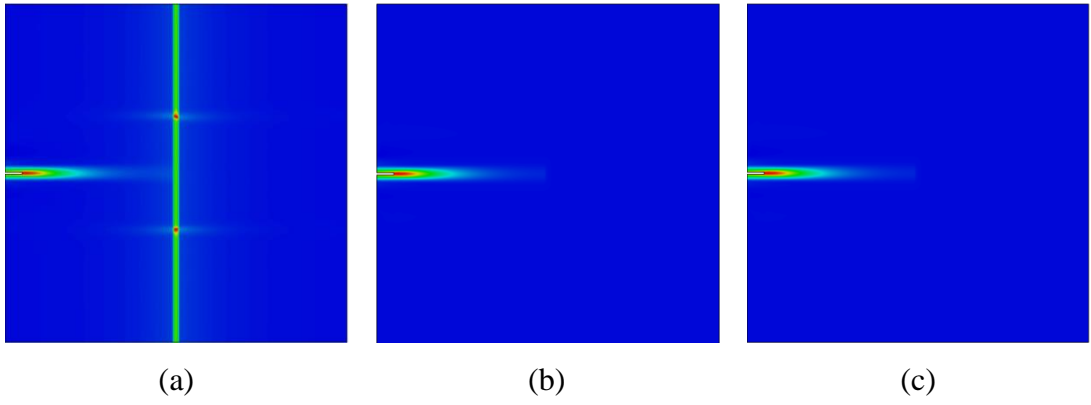

(a)            (b)            (c)

(2) Crack initial propagation stage

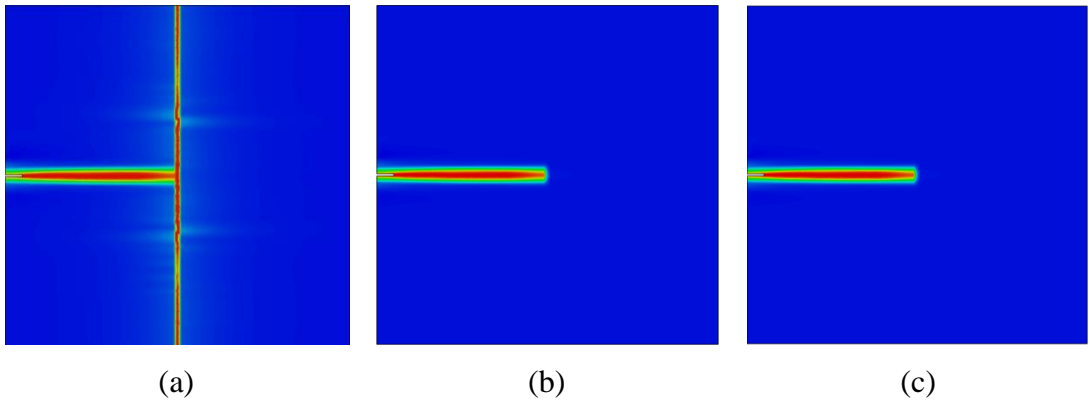

(a)            (b)            (c)

(3) Crack reaching interface stage

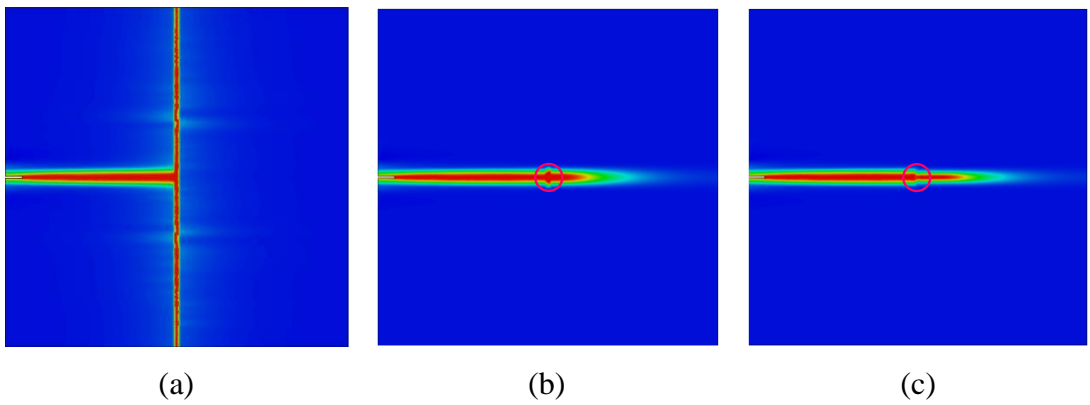

(a)            (b)            (c)

(4) Crack passing through the interface stage

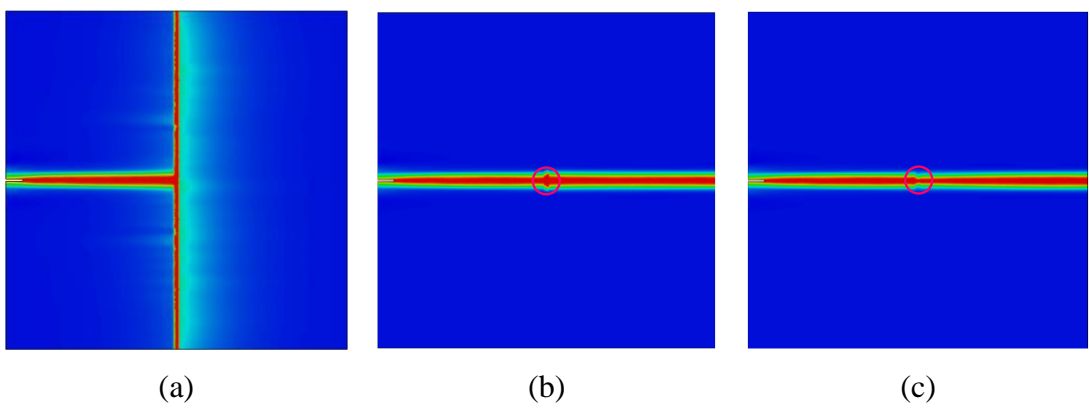

(a)            (b)            (c)

(5) Final failure stage

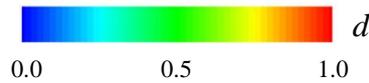

**Fig. 9. Crack evolution process: (a) simulation results of the weak interface; (b) strong interface 1 and (c) strong interface 2**

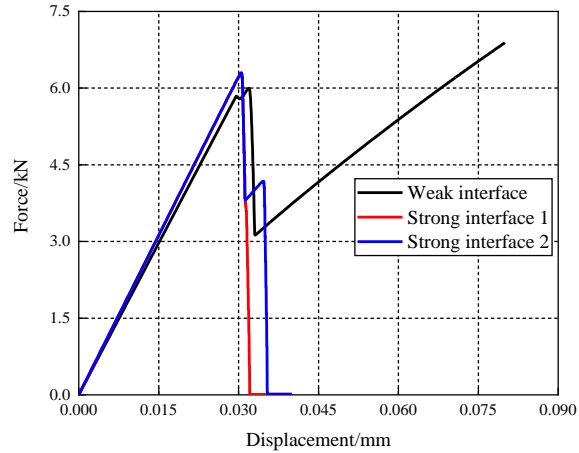

**Fig. 10. Load-displacement curve**

From the crack evolution results in Fig. 9, at the beginning of damage, the damage evolution process of the weak interface and strong interface is the same, as shown in Fig. 9(1). As the load continues to increase, the crack is initiates, and the damage evolution of the strong and weak interface begins to change. At the weak interface, because the fracture toughness of the interface material is weak, damage is initiated at the interface when the crack begins to expand. However, at the strong interface, due to the high fracture toughness of the interface material, there is no damage at the interface when the crack begins to propagate, as shown in Fig. 9(2). When the crack begins to reach the interface, the weak interface is partially destroyed, while the strong interface is not destroyed due to its high fracture toughness, as shown in Fig. 9(3). As the load continues to increase, the interface continues to break at the weak interface, but the crack does not pass through the interface and continue to propagate to the next layer of material. However, at the strong interface, the crack propagates rapidly through the interface to the next layer of material, and there is no bifurcation at the interface. At the stronger interface 2, the material shows stronger brittleness when the crack passes through the interface. Compared with strong interface 1, the damage formed at the interface is smaller, as shown in Fig. 9(4). When the load continues to increase, at the weak interface, after the interface is completely destroyed, the next layer of material begins to damage, but it has not yet been destroyed. At the strong interface, the crack has penetrated through the next layer of material, completely destroying the whole material, as shown in Fig. 9(5). It can be seen from the whole failure process that the crack propagates to the weak interface and bifurcates, which can delay the failure process of the material by consuming part of the energy, making the material show stronger fracture toughness. However, at the strong interface, the material shows brittleness, and the crack directly passes through the interface so that the interface does not play a role in delaying the failure process.

It can be seen from the load-displacement curve in Fig. 10 that in the strong interface material, the material failure process shows stronger brittleness, and the force load drops rapidly after reaching the maximum position, causing material failure. At the same time, because the fracture toughness of strong interface 2 is higher, more energy is consumed than that of strong interface 1 when failure occurs. Therefore, in the process of failure, there is a process of increasing and then decreasing, as shown by the solid blue line in Fig. 10. In the weak interface material, the interface consumes more energy, which delays the damage attenuation process, improves the failure strength of the material, and makes the material show stronger toughness.

### 4.2 Analysis of different critical energy release rates

In this section, the PFM with an effective critical energy release rate is used to predict the ultimate failure strength of composite materials. The material parameters are shown in Table 3, the model parameters are length L=80 mm, width W=18 mm, and circular hole radius $R$=2.5 mm, and the model is shown in Fig. 11. The angle between the fibre direction and vertical direction is $\alpha$ to simulate the damage of unidirectional composite plates with different fibre directions under tensile load, $\alpha$={0°, 15°, 30°, 45°, 60°}. The simulation results of cracks are shown in Fig. 12, and the final failure stress of the material is shown in Fig. 13.

**Table 3 Material parameters**

| Material | Paramter | Value | Unit |
|---|---|---|---|
|  | $E_{11}$ | 26.5 | GPa |
|  | $E_{22}$ | 2.6 | GPa |
| Property | $G_{12}$ | 1.3 | GPa |
|  | $v_{12}$ | 0.35 | - |
|  | $G_I$ | 0.622 | N/mm |

| | $G_{II}$ | 0.472 | N/mm |
|---|---|---|---|

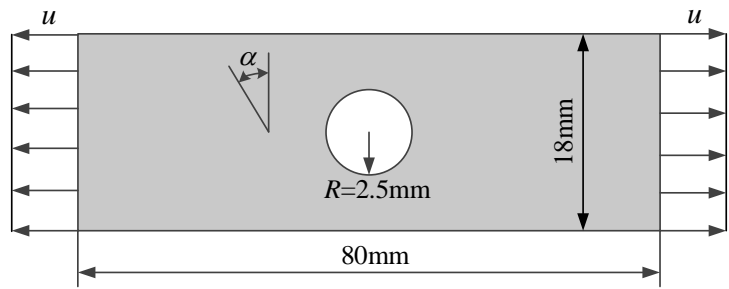

**Fig. 11. Geometric model**

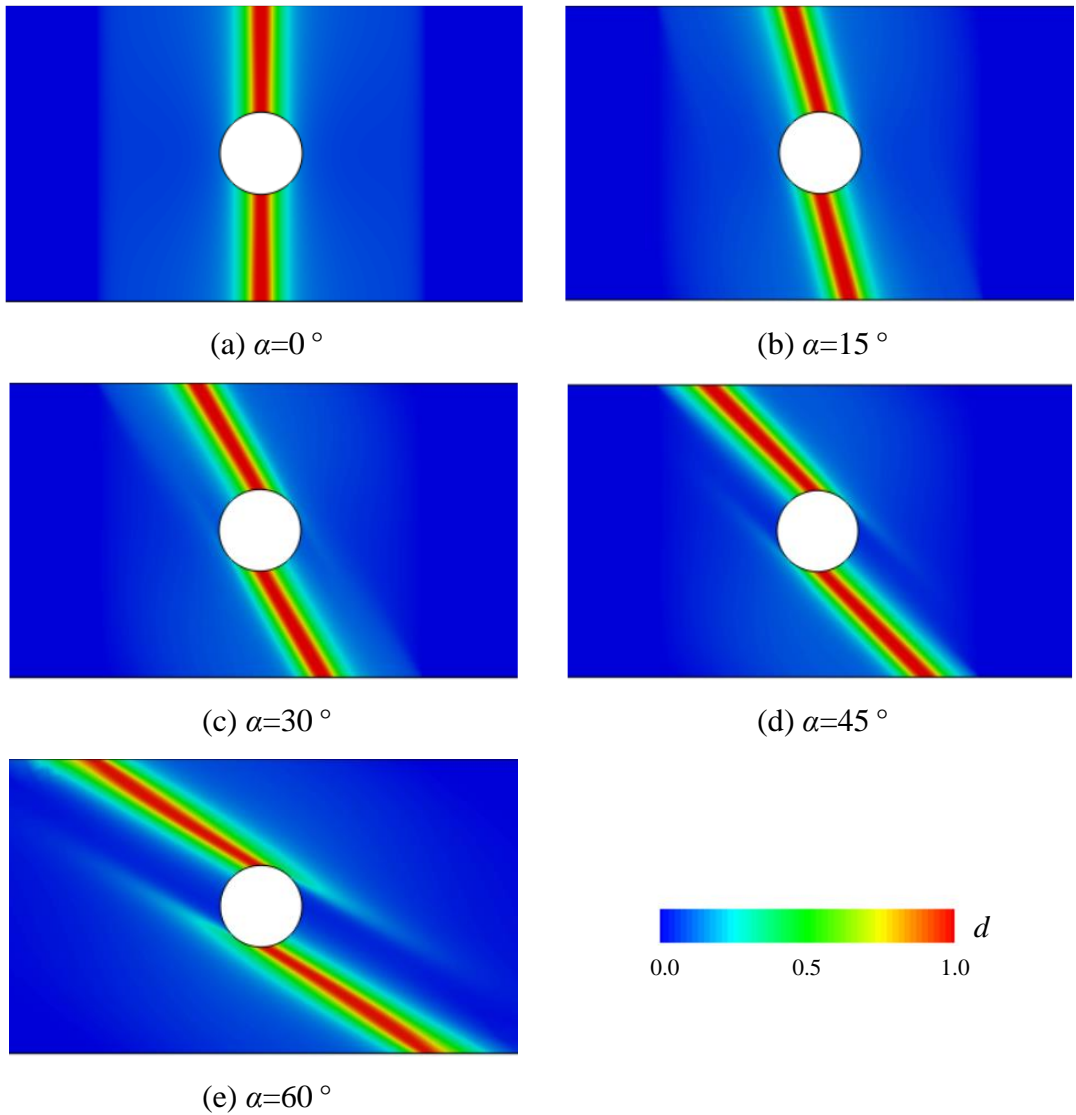

(a) $\alpha=0°$

(b) $\alpha=15°$

(c) $\alpha=30°$

(d) $\alpha=45°$

(e) $\alpha=60°$

**Fig. 12. Crack propagation results**

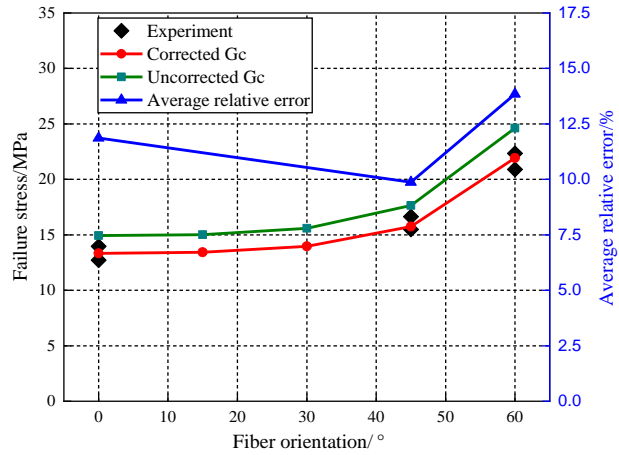

**Fig. 13. The failure stress of plates with holes in different fibre directions**

Fig. 12 shows that the macrocracks propagate along the fibre direction, causing fracture failure of the material. This is because fracture failure of the matrix occurs in this direction. Due to the high failure strength of the fibre, the overall fracture does not occur, and the energy required for the crack propagation of the material is less, so the damage evolves along the fibre direction and forms macrocracks, which is consistent with the theoretical analysis, and the simulation results are also consistent with the experimental observation in reference [56].

The simulation results of different critical energy release rates Gc and experimental results [57, 58] are shown in Fig. 13. It can be seen from the figure that in the case of $α≤30°$, the variation range of the simulated failure stress is very small because the angle between the fibre direction and the load direction is large; especially at α=0°, the angle between the fibre direction and the load direction is 90°. Additionally, the main structure of the force is the matrix at this time, the matrix consists of isotropic material, and the main failure mode is matrix failure. Therefore, within this angle range, the simulated failure stress has a small range of variation. In the case of $α>30°$, the angle between the fibre direction and the load direction gradually decreases, and the fibre gradually becomes a load-bearing structure, so the matrix and fibre begin to fail together. The greater the value of α is, the more load the fibre bears, and the load-bearing capacity of the fibre is stronger. Therefore, within this angle range, the failure stress begins to change significantly with the fibre angle; that is, the failure stress increases with increasing α.

At the same time, from the average relative error between the failure stress and the experimental value in Fig. 13, the error between the simulated failure stress and the experimental value using the uncorrected critical energy release rate Gc is large.

Especially when α=60°, the average relative error between the simulated result and the experimental value is approximately 13.8%. After correcting Gc, the failure stress is reduced; at different fibre angles, the simulation results are in good agreement with the experimental values, which verifies the accuracy of the correction equation of the critical energy release rate (44).

**4.3 Failure stress analysis of different hole shapes**

In this section, the PFM is used to predict the failure strength of the composite plate with different hole shapes. The material parameters are shown in Table 3. The model parameters are length L=80 mm and width W=18 mm. The area of different holes in the model is the same to ensure the same bearing area and eliminate the influence of the bearing area on the results. The angle between the fibre direction and horizontal direction is α=90°, and the model is shown in Fig. 14. The simulation results of crack propagation are shown in Fig. 15, the approximate analytical solution is the ratio of the circumferential stress at the edge of the hole to the load, and FEM is the simulated hole circumferential stress. The force-displacement curve of the ultimate failure of the material is shown in Fig. 16.

**Table 4 Material parameters**

| Material | Paramter | Value | Unit |
|---|---|---|---|
|  | $E_{11}$ | 47.4 | GPa |
|  | $E_{22}$ | 16.2 | GPa |
| Property | $G_{12}$ | 7 | GPa |
|  | $v_{12}$ | 0.26 | - |
|  | $G^{(*)}$ | 0.622 | N/mm |

(*) Due to the lack of material parameters, the critical energy release rate is assumed.

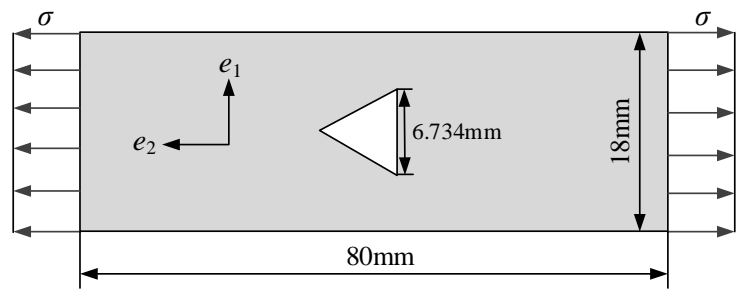

(a) Triangular hole

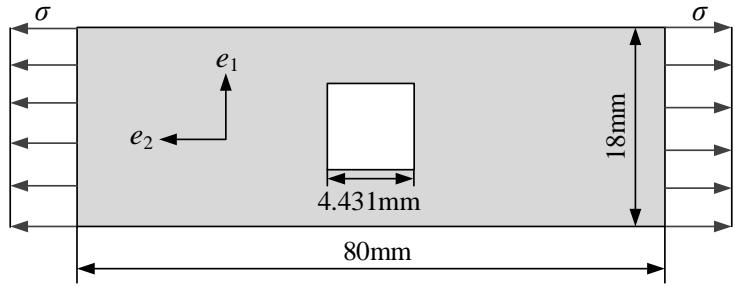

(b) Square hole

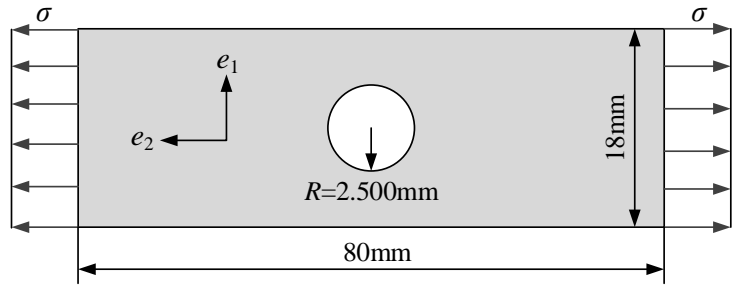

(c) Circular hole
**Fig. 14. Composite plates with different holes**

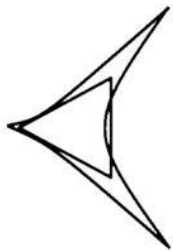 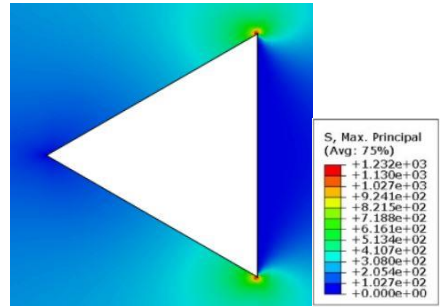

(a)Approximate analytical solution[59]　　　　(b)FEM solution

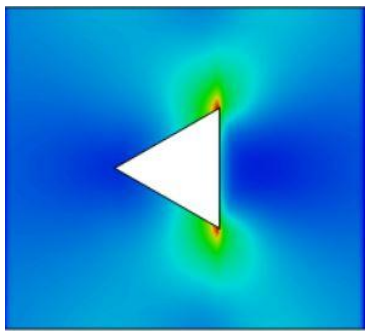 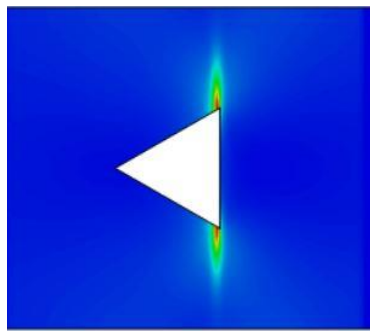 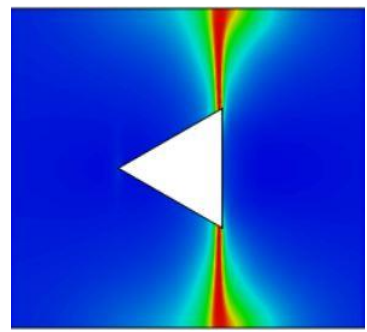

(c)u=0.002mm　　　　(d)u=0.152mm　　　　(e)u=0.166mm

(1) Crack propagation process of a triangular hole

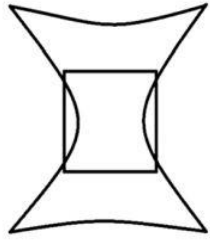
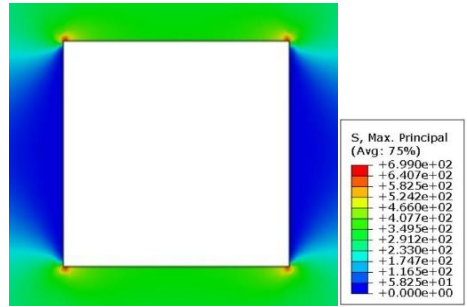

(a)Approximate analytical solution[59]  (b)FEM solution

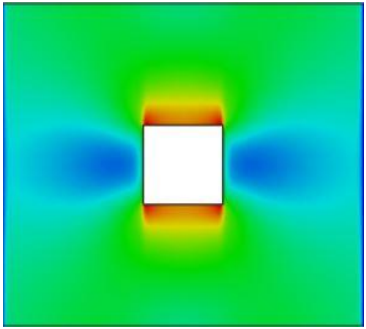
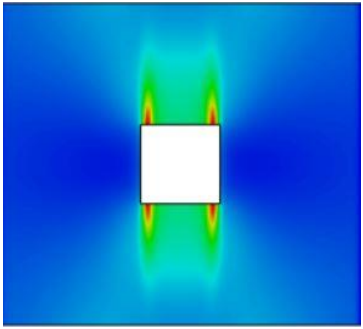
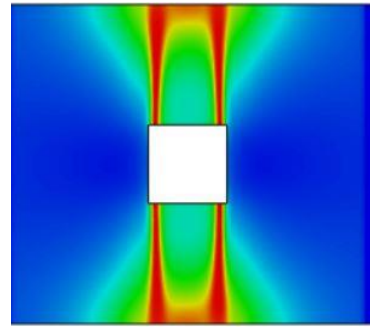

(c)u=0.002mm  (d)u=0.220mm  (e)u=0.234mm

(2) Crack propagation process of a square hole

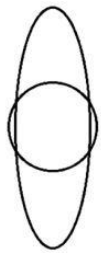
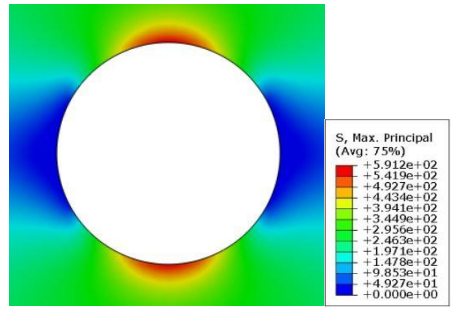

(a)Approximate analytical solution[59]  (b)FEM solution

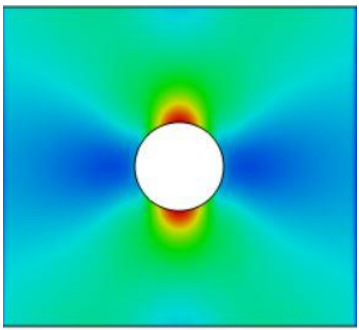
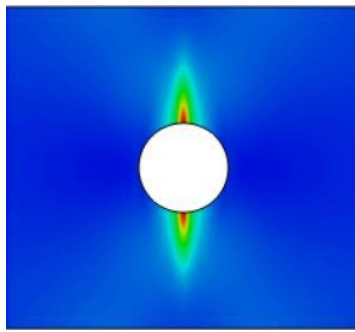
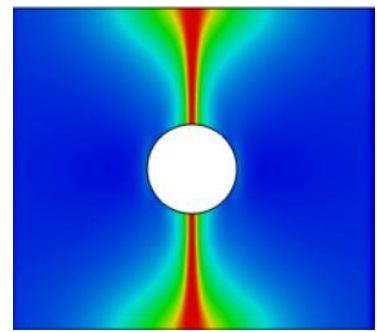

(c)u=0.002mm  (d)u=0.183mm  (e)u=0.198mm

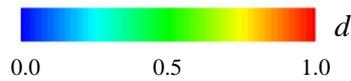

(3) Crack propagation process of a circular hole
**Fig. 15. Crack propagation process of different holes**

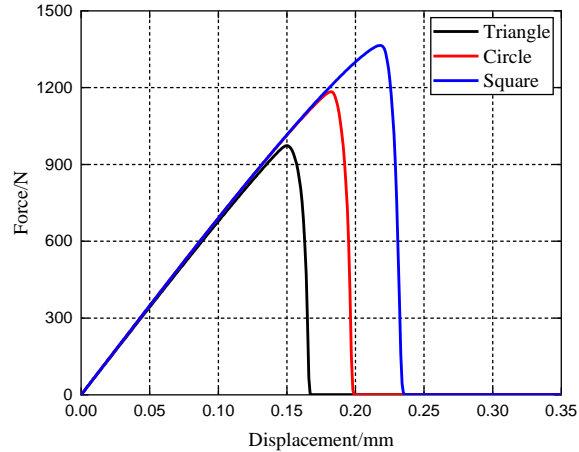

**Fig. 16. Load-displacement curves of different holes**

The Fig. 15 shows the stress distribution of composite plates with different hole shapes, as well as the change process of the initial damage stage, the initial crack propagation stage and the final failure stage of the material. It can be seen from Fig. 15(ia) and (ib) that stress concentration is prone to occur at sharp corners, and the simulation results are consistent with the analytical solution[59]. After comparison, it can be found that the stress concentration of triangular holes and circular holes only occurs on the upper and lower edges, while the stress concentration of square holes occurs at the four edge positions of the hole, and the stress concentration near the triangular hole is the most serious, but as the hole gradually tends to circular, under the same load, the stress concentration value of the hole is decreasing. Fig. 15(ic) shows the initial damage results of different hole types. It can be found that as the hole type gradually becomes circular, the initial damage concentration is gradually weakening. For example, in triangular and square holes, the damage is concentrated near the hole. In the area, but in the circular hole, the initial damage area is larger. This is because as the hole tends to be circular, the effect of stress concentration continues to weaken. Fig. 15(id) shows the results of different hole types of cracks starting to grow. The growth results are related to the hole type and force range. Due to the serious stress concentration near the triangle, the cracks begin to grow when the displacement u is small (Fig. 15(1d)), although the stress concentration near the square hole is smaller than that of the triangular hole, it is larger than that of circular holes. However, because the four positions of the upper and lower sides of the square hole are loaded, the total bearing capacity of square hole is larger than that of the triangular hole and the square hole, and the displacement $u$ at the beginning of crack propagation is larger (Fig. 15(2d), Fig. 15(3d)). Observing Fig. 15(ib) and Fig. 15(id), we can find that the location where the crack starts to grow is the location where the stress concentration occurs. At this location, the stress first reaches the destruction limit of the material, so damage occurs first, causing the material to fail. Comparing the displacement of different holes during crack propagation in Fig. 15(ib), $u_{triangle} < u_{circle} < u_{square}$ is obtained. Fig. 15(ie) show the results of crack propagation to the boundary of different holes, the crack propagation results of Fig. 15(1e) and Fig. 15(3e) are approximately similar, since damage and crack propagation only occurred near the upper hole in the triangular hole and circular hole, only one crack appeared near the upper hole. However, the square hole are loaded in four positions, so there are four cracks. Because the material properties begin to decline when the crack begins to propagate in the material, the displacement at the end of failure of materials with different holes is the same as that at the beginning of crack propagation, namely, $u_{triangle} < u_{circle} < u_{square}$.

It can be seen from Fig. 16 that the final bearing capacity is $F_{triangle} < F_{circle} < F_{square}$. For triangular and circular holes, from the beginning of the damage to the final failure process, the cracks start to expand from the edges of the upper and lower holes, causing the material to start to form cracks, and finally cause the failure of the entire material. However, the stress concentration of triangular holes is more serious than that of circular holes, so the triangle hole is damaged earlier than the circular hole, so the final failure load of the triangle hole is smaller than the circular hole. Although the stress concentration of square holes is more serious than that of circular holes, there are more stress concentration positions than circular holes, which results in more load bearing positions than circular holes at the time of failure, so the final bearing capacity is greater than that of circular holes.

## 5 Conclusion

In this paper, the effects of the interface properties, effective critical energy release rate, fibre orientation and hole shapes on the failure strength of composites are studied by using the anisotropic PFM, and the conclusions are as follows:

(1) For different interface properties of composite materials, the strong interface materials show brittleness during failure, while the weak interface materials show toughness during failure, which can improve the failure strength of materials.

(2) When a composite material fails, the ultimate failure stress increases with decreasing load direction and fibre direction. Due to the influence of the parameters $\alpha$ in the structure tensor on the failure stress, the existing PFM makes larger simulation results. By introducing the effective critical energy release rate, the fracture toughness of the material is reduced, and the simulation results are in good agreement with the experimental results.

(3) The bearing capacity of the composite with holes is related not only to the shape of the holes but also to the location of the concentrated stress and the number of locations where the holes bear the load. The greater the number of loading positions near the hole is, the stronger the bearing capacity is. Of the triangular, rhombic, square, hexagon, elliptical and circular holes studied in this paper, the square and elliptical holes demonstrate stronger bearing capacity.

## Declaration of Competing Interest

The authors declare that they have no known competing financial interests or personal relationships that could have appeared to influence the work reported in this paper.